\documentclass[reprint,superscriptaddress,amsmath,amssymb,aps,pra]{revtex4-2}
\usepackage[T1]{fontenc}
\usepackage{physics}
\usepackage[utf8]{inputenc}
\usepackage[english]{babel}
\usepackage{graphicx}     
\usepackage{hyperref}
\usepackage{xcolor}
\usepackage{soul}
\usepackage{float}
\usepackage[normalem]{ulem}

\hypersetup{colorlinks=true,citecolor={blue},linkcolor={blue},urlcolor={blue}}

\newcommand{\pd}[1]{{\begin{color}[rgb]{0.5,0,1}{#1}\end{color}}}

\begin{document}
\title{Quantum reservoir computing with classical and nonclassical states in an integrated optical circuit
}

\author{S.~Świerczewski} 
\affiliation{Center for Quantum Enabled-Computing, Center for Theoretical Physics of the Polish Academy of Sciences, Al. Lotników 32/46, 02-668 Warsaw, Poland}
\author{W. Verstraelen}
\affiliation{Division of Physics and Applied Physics, School of Physical and Mathematical Sciences, Nanyang Technological University, Singapore, Singapore} 
\affiliation{Majulab, International Joint Research Unit UMI 3654, CNRS, Université Côte d’Azur, Sorbonne Université, National University of Singapore, Nanyang Technological University, Singapore, Singapore 117543}
\author{P. Deuar}
\affiliation{Institute of Physics, Polish Academy of Sciences, Aleja Lotnik\'ow 32/46, PL-02-668 Warsaw, Poland}
\author{T. C. H. Liew} \affiliation{Division of Physics and Applied Physics, School of Physical and Mathematical Sciences, Nanyang Technological University, Singapore, Singapore} 
\affiliation{Majulab, International Joint Research Unit UMI 3654, CNRS, Université Côte d’Azur, Sorbonne Université, National University of Singapore, Nanyang Technological University, Singapore, Singapore 117543}
\author{A.~Opala}
\affiliation{Institute of Experimental Physics, Faculty of Physics, University of Warsaw, ul. Pasteura 5, PL-02-093` Warsaw, Poland}
\affiliation{Institute of Physics, Polish Academy of Sciences, Aleja Lotnik\'ow 32/46, PL-02-668 Warsaw, Poland}
\author{M.~Matuszewski}
\affiliation{Center for Quantum Enabled-Computing, Center for Theoretical Physics of the Polish Academy of Sciences, Al. Lotników 32/46, 02-668 Warsaw, Poland}
\affiliation{Institute of Physics, Polish Academy of Sciences, Aleja Lotnik\'ow 32/46, PL-02-668 Warsaw, Poland}

\begin{abstract}
Quantum reservoir computing (QRC) is a hardware-implementation-friendly quantum neural network scheme with minimal physical system requirements and a proven advantage over classical counterparts. We use an extension of the positive-$\mathcal{P}$ phase space method to efficiently simulate a bosonic, linear silicon-chip based QRC system excited with a single nonclassical state, a ``kitten'' state. In combination with input-encoding coherent states, our method allows to obtain exact results for all correlation functions without Hilbert space cutoff. Surprisingly, we find that such a setting - where the only ``quantumness'' derives from a single input mode, is sufficient to obtain significant (over 9-fold) reduction of classification error over the classical counterpart. Our work provides a promising direction toward efficient quantum computation with accessible optical hardware.
\end{abstract}
\maketitle

\section{Introduction}

Quantum technologies have emerged as a useful tool for information processing, offering advantages in computing, sensing, and simulation~\cite{arute2019quantum,degen2017quantum,georgescu2014quantum}. Within this landscape, Quantum Neural Networks (QNNs) have attracted significant attention due to their potential to process information in ways that classical counterparts cannot efficiently emulate~\cite{schuld2014quest,beer2020training}. A particularly interesting paradigm of QNNs is QRC~\cite{fujii2017harnessing,ghosh2019quantum,ghosh2021quantum}. While conventional QNNs require complex unitary gates, QRC is based on non-linear dynamics of a fixed reservoir quantum system to map inputs into a high-dimensional feature space, and requires training only the final linear output layer, often implemented in software. This architecture reduces the training overhead and improves convergence, making it particularly attractive for near-term noisy quantum devices.

On the fundamental level, the use of non-classical quantum states is required to achieve a computational advantage over classical systems. However, the theoretical description and numerical simulation of such states and their dynamics face a bottleneck: the exponential scaling of the Hilbert space dimension with the system size. For systems with a large number of modes, exact computation becomes computationally intractable, except for specific systems and inputs such as Gaussian states and linear dynamics~\cite{weedbrook2012gaussian}. To address this issue, approximate methods have been developed, including cumulant expansion \cite{Casteels2016,WVcorrelations} phase space methods based on the Wigner or positive-$\mathcal{P}$ representations~\cite{gardiner2004quantum}, and tensor network methods~\cite{orus2014practical,van2025optimizing}, which are powerful but can struggle with entangled evolution in two or more dimensions. Achieving an efficient numerical description, which often means an exponential reduction in numerical complexity, remains a critical challenge. Currently, there is no universal paradigm or method suitable for every system configuration and all parameter regimes.

In this work, we show that an extension of the Positive-$\mathcal{P}$ representation~\cite{Drummond_1980} is ideal for describing dynamical systems initialized with specific classical and non-classical states. We consider inputs consisting of coherent states, cat states, and their small-amplitude variants known as ``kitten'' states. The extension we use involves allowing for complex amplitudes in phase-space trajectories, effectively rendering the method a ``generalized-$\mathcal{P}$ representation.'' While theoretical foundations of such generalized-$\mathcal{P}$ representations have been established in quantum optics and atom optics~\cite{Drummond_1980,Deuar02,DeuarPhD,Teh}, here we apply this formalism to the practical setting of a quantum neural network implementable in a silicon photonics chip~\cite{tait2017neuromorphic,shen2017deep}.

\begin{figure*}[bht!]
\centering

\includegraphics[width=1\linewidth]{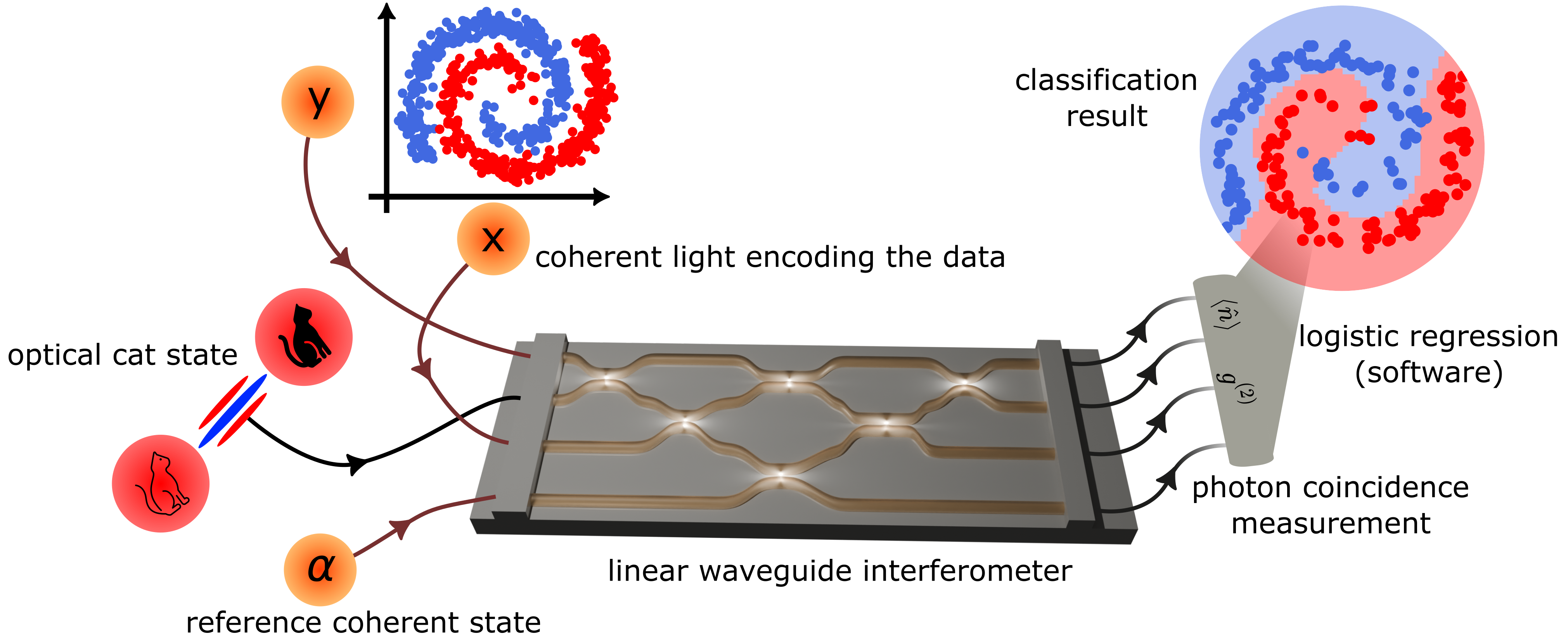}

\caption{Schematic representation of QRC using coherent input states interacting with a non-classical state in a linear waveguide interferometer. The data to be classified, which in this case consists of two spirals parametrized by the coordinates x and y, is encoded in phases or amplitudes of coherent inputs. Additionally, a non-classical state (in our case the optical ``cat'' or ``kitten'' state) and a coherent field with a fixed amplitude and phase are injected. The interferometer produces multi-mode entanglement even in the case of linear coupling. At the output of the network, average occupations $\langle \hat{n}_{i}\rangle$ and multi-mode correlations $g^{(2)}_{ij}$ are measured, effectively performing a nonlinear feature extraction. These features are used in the output layer for software logistic regression. During training, only the output weights are modified and the physical system remains intact.}
    \label{fig:fig1}
\end{figure*}

We address a question of practical importance in the design of quantum networks: is the injection of a single non-classical input state, while encoding all other data in classical states, sufficient to achieve a quantum improvement in prediction accuracy? We answer this in the affirmative for the specific case where the non-classical state is a ``kitten'' state, and the remaining inputs are coherent states (See Fig.~\ref{fig:fig1}). We demonstrate a qualitative improvement in QRC performance compared to purely classical input baseline. By employing the generalized-$\mathcal{P}$ phase space description, we can obtain exact results independent of the number of modes, occupations or the geometric complexity of the photonic circuit. This result suggests a scalable pathway for designing large-scale photonic quantum reservoir networks with minimal requirements on both the physical device and the input states.

\section{Generalized-$\mathcal{P}$ description and cat states}

Quantum optics often involves multimode electromagnetic fields, each described by bosonic annihilation and creation operators obeying canonical commutation relations \cite{WallsMilburn2008}. As the number of modes increases, the dimension of the associated Hilbert space grows exponentially, making direct operator-based calculations computationally inefficient. To address this challenge, phase-space representations such as the Wigner \cite{Wigner_1932}, truncated Wigner \cite{Steel98,Sinatra02}, Glauber–$\mathcal{P}$ \cite{Glauber}, positive-$\mathcal{P}$ \cite{Drummond_1980,Deuar_2021}, and Husimi–Q \cite{Husimi} functions are widely used to model multimode quantum systems.

In our work, we study a network of coupled single-mode waveguides in which various quantum states—including states with high average occupation—propagate and interact across different modes. Such a network also exhibits exponential Hilbert-space growth, necessitating an efficient phase-space representation. Because the initial states in our proposal include both coherent states and their superpositions (commonly referred to as “cat’’ or “kitten’’ states for superpositions with amplitudes $|\alpha| \sim 1$), we must select a representation capable of treating both types of states accurately. While coherent states are easily modeled due to their Gaussian representation in phase space and close correspondence with classical field descriptions, superpositions of distinct coherent states pose significant challenges. For example, the Glauber–$\mathcal{P}$ distribution for a single coherent state is a Dirac delta distribution, whereas for a superposition of two coherent states it becomes a highly singular object involving derivatives of delta functions.

To accurately simulate both coherent states and coherent-state superpositions, we therefore employ the generalized-$\mathcal{P}$ representation, which introduces two independent phase-space variables instead of the single complex variable used in the Glauber–$\mathcal{P}$ representation. This enlarged phase space regularizes the representation of nonclassical states—such as cat and kitten states—making it well suited for our multimode simulations.

Here, we consider a generalized-$\mathcal{P}$ distribution~\cite{Drummond_1980}, which allows one to represent an arbitrary density matrix according to
\begin{equation}
\hat{\rho}
=\int d^{2}\alpha d^{2}\tilde{\alpha}\ 
\mathcal{P}(\alpha,\tilde{\alpha}^{*})
\hat{\Lambda}(\alpha,\tilde{\alpha}^{*}),
\label{eq:dminP}
\end{equation}
where the kernel operators $\hat{\Lambda}$ are non-Hermitian coherent-state projectors with unit trace, defined as
\begin{equation}
\hat{\Lambda}(\alpha,\tilde{\alpha}^*)
=\frac{\ket{\alpha}\bra{\tilde{\alpha}}}
{\braket{\tilde{\alpha}}{\alpha}},
\qquad
\mathrm{Tr}\big[\hat{\Lambda}\big]=1.
\label{lambda}
\end{equation}
If one restricts the kernel to its diagonal elements, $\alpha=\tilde{\alpha}$, the representation reduces to the standard Glauber–$\mathcal{P}$ distribution.
An arbitrary cat state formed as a superposition of two coherent states with opposite amplitudes can be written as
\begin{equation}
\ket{\psi}
=\frac{1}{\sqrt{\mathcal{N}}}\left(
a\ket{\beta}
+be^{\mathrm{i}\theta}\ket{-\beta}
\right),
\end{equation}
where $a$ and $b$ are real coefficients satisfying $a^{2}+b^{2}=1$, and $\mathcal{N}$ is the normalization factor.
This state can be represented using the generalized-$\mathcal{P}$ function of the form
\begin{align}
\mathcal{P}(\alpha,\tilde{\alpha}^*)
&=\frac{1}{\mathcal{N}}\Big[
a^{2}
\delta^{(2)}(\alpha-\beta)
\delta^{(2)}(\tilde{\alpha}-\beta)
 \nonumber \\
&\quad
+abe^{i\theta}
\langle \beta | -\beta\rangle
\delta^{(2)}(\alpha+\beta)
\delta^{(2)}(\tilde{\alpha}-\beta)
\nonumber \\
&\quad
+abe^{-i\theta}
\langle -\beta | \beta\rangle
\delta^{(2)}(\alpha-\beta)
\delta^{(2)}(\tilde{\alpha}+\beta)
\nonumber \\
&\quad
+b^{2}
\delta^{(2)}(\alpha+\beta)
\delta^{(2)}(\tilde{\alpha}+\beta)
\Big].\label{eq:catP}
\end{align}
A direct substitution of Eq.~\eqref{eq:catP} into Eq.~\eqref{eq:dminP} verifies that the resulting density matrix coincides with the cat-state density operator. The form of this distribution is similar to the positive-$\mathcal{P}$ but complex-valued or complex-weighted (hence a kind of gauge-$\mathcal{P}$ distribution \cite{Deuar02} in principle), which requires special treatment.

\begin{figure*}[bht!]
    \centering
    \includegraphics[width=1\linewidth]{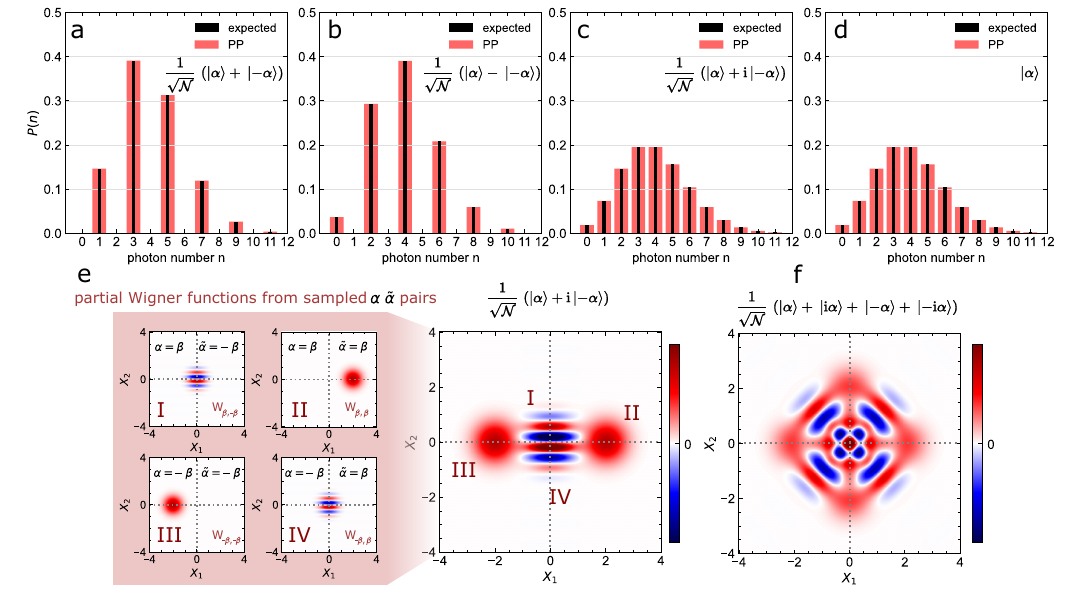}
    \caption{Panels \textbf{a - d} present the photon-number probability distribution calculated from the weighted complex-$\mathcal{P}$ distribution for different superpositions of the $\ket{\alpha}$ and $\ket{-\alpha}$ coherent states with coherent amplitude $\alpha = 2$. Panel \textbf{a} presents the photon number distribution for an ``even'' cat state built from even Fock states only and panel \textbf{b} for an ``odd'' cat state built from odd Fock states. The photon number distributions in panels \textbf{c} and \textbf{d} are identical, although the distribution on panel \textbf{c} corresponds to a cat state and the one on panel \textbf{d} to a coherent state. Panel \textbf{e} depicts the Wigner function for a cat state, together with the partial Wigner functions (W$_{\gamma,\delta}$) calculated from the compact complex-$\mathcal{P}$ distribution by choosing coherent states $\gamma$ and $\delta$. Panel \textbf{f} depicts a higher-order cat state being a superposition of four coherent states located in the vertices of a square centered at zero in the phase space.}
    \label{fig:fig2}
\end{figure*}

To evaluate expectation values of observables $\hat{O}$ one may borrow techniques from the positive-$\mathcal{P}$ formalism \cite{Teh}.
If the complex generalized distribution $\mathcal{P}$ can be decomposed into a sum of positive distributions $\mathcal{P}^{+}_{\nu}$ with complex weights $\mathrm{w}_{\nu}$ (which is the case for coherent-state superpositions such as cat states),
\begin{equation}
\mathcal{P}(\alpha,\tilde{\alpha}^{*})
= \sum_{\nu} \mathrm{w}_{\nu}\mathcal{P}^{+}_{\nu}(\alpha,\tilde{\alpha}^{*}),
\end{equation}
then expectation values can be computed via weighted averages over the positive components, while keeping track of the complex coefficients $\mathrm{w}_{\nu}$.
Then one can perform the following calculation of the expected value of observable $\hat{O}$
\pd{\cite{DeuarPhD}}.
\begin{align}
\left\langle \hat{O} \right \rangle
    &= \mathrm{Tr}(\hat{O}\hat{\rho})  \nonumber\\
    &= \int {d^2}{\alpha}\,{d^2}\tilde{{\alpha}}\ \mathcal{P}({\alpha},{\tilde{\alpha}^*})\mathrm{Tr}(\hat{O}\hat{\Lambda}). \nonumber\\
    &= \int {d^2}{\alpha}\,{d^2}\tilde{{\alpha}}\,\sum_{\nu} \mathrm{w}_{\nu} \mathcal{P}^{+}_{\nu}({\alpha},{\tilde{\alpha}^*})\mathrm{Tr}(\hat{O}\hat{\Lambda}). \nonumber\\
    &= \sum_{\nu} \mathrm{w}_{\nu}\int {d^2}{\alpha}\,{d^2}\tilde{{\alpha}}\, \mathcal{P}^{+}_{\nu}({\alpha},{\tilde{\alpha}^*})\mathrm{Tr}(\hat{O}\hat{\Lambda}). \nonumber\\
    &= \sum_{\nu} \mathrm{w}_{\nu}\lim_{S\to\infty} \langle \mathrm{Tr}(\hat{O}\hat{\Lambda}) \rangle_{S,\mathcal{P}^{+}_{\nu}},\label{obs}
\end{align}
where $\mathrm{Tr}[\hat{O},\hat{\Lambda}]$ due to the properties of $\hat{\Lambda}$ yields a c-number expression containing $\alpha$ and $\tilde{{\alpha}}$. The last step of the derivation reads that an average over a stochastic ensemble of $S$ 
trajectories calculated by sampling the different $\mathcal{P}^{+}_{\nu}$ distributions is to be calculated , becoming ever more accurate as $S$ grows. This can be performed, because the $\mathcal{P}^{+}_{\nu}$ are chosen to be proper probability distributions. Conveniently, for delta-like distributions (as for the presented coherent state superposition) this means choosing only one pair of $\alpha$ and $\tilde{{\alpha}}$. This ``weighted'' method~\cite{Teh} has proven to be useful and was used to obtain all results presented in this manuscript. Fig.~\ref{fig:fig2} \textbf{a - d} demonstrates 
the utilization of the weighted probability method to calculate the photon number probability distribution for four different states, including cat states with negative (odd state on panel \textbf{b}) and complex (state on panel \textbf{c}) weights. The results coincide with the analytically expected values. These probability distributions were calculated by finding the expectation values of the $n$-th photon state operator $\ket{n}\bra{n}$ on these states.

\subsection{Wigner function from the generalized-$\mathcal{P}$ distribution}
The generalized-$\mathcal{P}$ representation is a four-dimensional representation and therefore impractical 
to efficiently plot and visualize. A more intuitive phase-space representation is the Wigner function representation. Knowing how to draw the Wigner function for a given $\mathcal{P}$ representation would be useful, as we could monitor the evolution of the quantum system not only by calculating observables but also by looking at the phase-space portrait of the quantum state. For a positive-$\mathcal{P}$ distribution, where the fields $\alpha$ and $\tilde{\alpha}$ are sampled from the $\mathcal{P}$ distribution, the Wigner function can be reconstructed by adding ``partial'' Wigner functions corresponding to a given sampled pair according to the following equation (see derivation in the Methods section \ref{Wigner})
\begin{align}
    &W\left(\xi, \xi^*\right)=  \nonumber\\ 
    &\lim_{S \xrightarrow{} \infty}\frac{2}{S \pi} \sum_{j=1}^S \operatorname{Re}\left\{\exp \left(-2\left(\xi^*-\tilde{\alpha}_j^*\right)\left(\xi-\alpha_j\right)\right)\right\},
    \label{WfromPP}
\end{align}
where \pd{$S$} is the number of samples taken from the $\mathcal{P}$ distribution. For the case of a  generalized-$\mathcal{P}$ distribution, like in Eq.~(\ref{obs}),
we need to multiply the ``partial'' Wigner functions by corresponding weights in the right way. Following Methods, this 
yields 
\begin{align}
W(\xi,\xi^*) &= \sum_{\nu}\lim_{S\to\infty}\frac{2}{S\pi}
\sum_{j=1}^S \Re\Big\{ \nonumber \\
&\mathrm{w}_{\nu}\,
\exp\big(-2(\xi^*-\tilde\alpha_{\nu,j}^*)(\xi-\alpha_{\nu,j})\big)\Big\}.
\end{align}
For the cat state, where the generalized-$\mathcal{P}$ distribution consists of Dirac delta functions, this expression simplifies due to choosing only one $(\alpha_{\nu}$,$\tilde{\alpha}_{\nu})$ pair per complex weight\\
\begin{align}
W(\xi,\xi^*) &= \sum_{\nu}\frac{2}{\pi}
 \Re\Big\{ \nonumber \\
&\mathrm{w}_{\nu}\,
\exp\big(-2(\xi^*-\tilde\alpha_{\nu}^*)(\xi-\alpha_{\nu})\big)\Big\},
\label{WfromPP}
\end{align}
where the pairs $(\alpha_{\nu}$,$\tilde{\alpha}_{\nu})$ are the centers of the delta functions in Eq.~\eqref{eq:catP}.

Figure~\ref{fig:fig2}\textbf{e} presents the ``reconstruction'' of the Wigner function of a cat state $\ket{\psi} \propto (\ket{\beta} + \mathrm{i}\ket{-\beta})$ with $\beta = 1$, in which case $\mathrm{w}_1=\mathrm{w}_4=1, \mathrm{w}_2=i,\mathrm{w}_3=-i$. Depending on the sampled pair $(\alpha_{\nu}$,$\tilde{\alpha}_{\nu})$, the partial Wigner function reconstructs a different part of the cat state. If the two variables are equal, the partial Wigner function is proportional to that of a coherent state, however where they are different, the interference fringes are reconstructed. This method can be applied to more complicated states, such as a superposition of four coherent states that are equidistant in the phase space (a higher-order cat state) yielding the Wigner function presented in Fig.~\ref{fig:fig2}\textbf{f}.

\section{Waveguide interferometer}

We consider an array of single-mode, lossless waveguides, where only neighboring waveguides are coupled, as shown in 
Fig.~\ref{fig:fig1}. We assume that the system is designed in such a way that the coupling between the $i$-th and $i$+1-th waveguide depends on the propagation variable $z$ and is described by the super-Gaussian function 
\begin{equation}
    J_{i,i+1}(z) = J\exp{-\left(\frac{(z-z_0)}{\sqrt{2}\sigma}\right)^{d}},
\end{equation}
where $J$ is the coupling strength, $z_{0}$ is center of the coupling region, $\sigma$ is the width, and $d$ is the super-Gaussian exponent (see Figure \ref{fig:fig3} \textbf{a}). This exponent must be even, and the larger the exponent, the ``steeper'' the slope of the coupling function, becoming a ``top-hat'' function for infinite $d$. We assume that ultrashort optical pulses propagate in the positive $z$ direction with no backscattering. The evolution of a state described by density matrix $\rho(t)$ is given by the Von-Neumann equation\\
\begin{equation}
    \frac{d}{dt}\hat{\rho}(t) = \frac{1}{\mathrm{i}\hbar}\left[\hat{H}(z), \hat{\rho}(t)\right], 
    \label{eq:vonneumann}
\end{equation}
where
\begin{align}
    \hat{H}(z) &= \sum_{i}\hat{H}_{i}(z), \\
    \hat{H}_{i}(z) & =\hbar\omega_i \hat{a}_i \hat{a}_i^{\dagger}+J_{i, i-1}(z)\left(\hat{a}_i^{\dagger} \hat{a}_{i-1}+\hat{a}_i \hat{a}_{i-1}^{\dagger}\right) \nonumber\\
    &+J_{i+1, i}(z)\left(\hat{a}_i^{\dagger} \hat{a}_{i+1}+\hat{a}_i \hat{a}_{i+1}^{\dagger}\right),
\end{align}
here $\omega_i$ is the on-site energy of the $i$-th waveguide mode and $\hat{a}_i$ and $\hat{a}_{i}^{\dagger}$ are respectively the $i$-th mode creation and annihilation operators.  If we now assume, that photons in each waveguide are propagating with a constant and equal group velocity $v_{g}$, we can perform a 
transformation of the Von-Neumann equation (\ref{eq:vonneumann}) from time to space dependence, by rewriting time as $t \xrightarrow{} z/v_{g}$. Now the rewritten evolution equation takes the form\\
\begin{equation}
    \frac{d}{dz}\hat{\rho}(z) = \frac{1}{\mathrm{i}\hbar v_{g}}\left[\hat{H}(z), \hat{\rho}(z)\right].
    \label{eq:vonneumann_x}
\end{equation}
This equation can (in principle) be solved by representing all operators in the Fock basis and then integrating using standard methods. The issue with this approach is that the dimension of the density matrix (and other operators in equation (\ref{eq:vonneumann_x})) grows as $\sim n_{\rm cutoff}^N$ where $n_{\rm cutoff}$ is the level of the Fock space truncation and $N$ is the number of modes. This exponential scaling makes this integration method inefficient and has led us to choosing the positive-$\mathcal{P}$ method as an alternative.

In the positive-$\mathcal{P}$ representation, the evolution equations for our system are deteministic and can be written as \\
\begin{subequations}
\begin{align}
& \frac{\partial \alpha_i}{\partial z}=\frac{\mathrm{i}}{\hbar v_{g}}\left[ \hbar\omega_i \alpha_i+ J_{i,i-1}(z) \alpha_{i-1} + J_{i,i+1}(z) \alpha_{i+1}\right], \\
& \frac{\partial \tilde{\alpha}_i}{\partial z}=\frac{\mathrm{i}}{\hbar v_{g}}\left[ \hbar\omega_i \tilde{\alpha}_i+  J_{i,i-1}(z) \tilde{\alpha}_{i-1} +  J_{i,i+1}(z) \tilde{\alpha}_{i+1}\right],
\end{align}
\end{subequations}
where $\alpha_{i}$ and $\tilde{\alpha}_{i}$ are the complex local coherent state amplitudes of the $i$-th mode. As there is no in-mode two-photon interaction here, no stochastic evolution term of the kind commonly seen in positive-$\mathcal{P}$ simulations appears.  In the positive-$\mathcal{P}$ formalism, instead of solving a matrix differential equation, a set of coupled differential equations is solved with the number of equations needed to be solved growing linearly with the number of waveguide modes. We used the exponential midpoint algorithm \cite{Deuar2021multitime} to solve this set of equations. We  assumed all frequencies $\omega_{i}$ to be equal and solved the equations in the frame rotating with this frequency. this is equivalent to setting all $\omega_i = 0$.

In all simulations we chose $v_{g}$ to resemble the group velocity of light in silicon at telecom wavelength ($\sim 1550$ nm) $v_{g} \sim c/n_{\mathrm{Si}} \approx 70 \, \mathrm{\mu m/ps}$~\cite{Dulkeith:06_group}. We  assumed the coupling strength $J_{ij}$ values to be of the order of magnitude of a few meV\cite{Nigro2022IntegratedQuantumPolaritons}. Based on experimentally realised photonic waveguide interferometers, we also keep the length of the waveguides in our simulation within a few centimeters \cite{Lenzini_waveguide}. Such a setup leads to the time needed to propagate through the waveguide array in the range of hundreds of picoseconds.

\subsection{Two-mode interferometer}
\begin{figure}[bht!]
    \centering
    \includegraphics[width=1\linewidth]{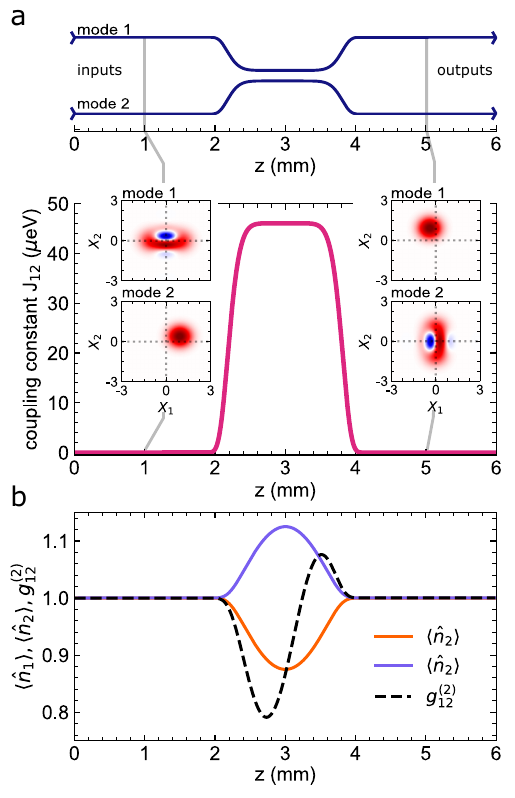}
    \caption{Panel \textbf{a} depicts a schematic representation of a two-mode waveguide interferometer together with the linear coupling strength between the two modes as a function of the waveguide length along the propagation direction. The sub-panels represent the partial single-mode Wigner functions of the states in each mode before entering (on the left) and after exiting (on the right) the coupling region, as calculated using Eq. (\ref{WfromPP}). One can see, that the kitten state has initially been in mode 1, after it is coupled to the second mode it is transferred to mode 2. Panel \textbf{b} shows the average photon number in both modes and the cross-mode $g_{12}^{(2)}=\langle\hat{a}^{\dagger}_1\hat{a}^{\dagger}_2\hat{a}_1\hat{a}_2\rangle / \langle \hat{n}_1\rangle \langle \hat{n}_2\rangle$ correlation function.  }
    \label{fig:fig3}
\end{figure}

We first study interference between a coherent state and a kitten state initially being in two different waveguide modes. The schematic representation of a two-mode interferometer, together with the input and output state Wigner functions and the evolution of average occupation and multimode correlations is presented in Fig.~\ref{fig:fig3}. Waveguide mode 1 was initialized in a kitten state $\ket{\psi} \propto \ket{\beta} + \mathrm{i}\ket{-\beta}$ with $\beta = 1$ and mode 2 in a coherent state with complex amplitude $\beta_{\mathrm{coh}} = e^{\mathrm{i}\pi/8}$. Throughout the evolution, the ``quantumness'' of the kitten state does not disappear as we assume no photon loss or other decoherence sources. However, it is distributed among both of the waveguide modes. If the coupling strength and coupling region length is chosen in a specific way, it is even possible for the kitten state to transfer to the second mode and this is the case shown in Fig.~\ref{fig:fig3}. It is worth noting that the second order cross-correlation function, as well as the average occupation of the two modes, change during propagation through the coupling region. All calculations presented in Figure ~\ref{fig:fig3} were done with the use of the weighted probability distributions and methods to represent the Wigner function presented in the previous section.

\begin{figure}[bht!]
    \centering
    \includegraphics[width=1\linewidth]{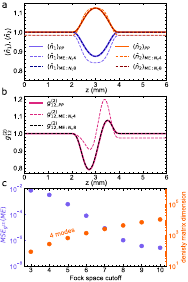}
    \caption{Panel \textbf{a} depicts the average occupation on each mode of a two-mode waveguide interferometer, where initially mode 1 was in a kitten state and mode 2 in a coherent state. The continuous line represents the result of the exact PP calculation, and the dashed lines present the result of solving the Von Neumann equation with a density matrix represented in the Fock space basis and Fock space cutoff 4 (light colored) and 8 (dark colored) lines. Panel \textbf{b} presents the comparison between the multi-mode correlation function calculated in the PP framework and using the von Neumann equation. Panel \textbf{c} presents the mean-squared-error ($MSE_{g^{(2)}}$) of calculating the $g^{(2)}_{12}$ function using the master equation (ME) integrated along the propagation direction for different levels of Fock space truncation. The right axis depicts the density matrix dimension of a 4-mode network (as studied in Fig \ref{fig:fig5}) assuming a given Fock space cutoff. This highlights the importance of finding a optimal cutoff to minimize the error but keeping the density matrix dimension within computational possibilities. The interferometer setup and input states have been set to the same as in the calculations presented in Fig.~\ref{fig:fig3}}
    \label{fig:fig4}
\end{figure}

To motivate the employment of the generalized-$\mathcal{P}$ method, we compare it with the standard practice of solving the density matrix evolution equation in the multimode Fock state basis. The first important observation is that the generalized-$\mathcal{P}$ simulations are exact. We do not need to apply any approximations, e.g.~resulting from Fock space truncation or sampling over stochastic trajectories. Both sampling the initial state as well as the evolution is deterministic. Fig.~\ref{fig:fig4} depicts the error of calculating the average occupation and the cross-mode $g^{(2)}$ correlation function depending on the Fock space cutoff. Moreover, the integrated mean squared error of calculating the correlation function using the Von Neumann equation (\ref{eq:vonneumann_x}) is plotted as a function of the Fock space cutoff. The Figure shows that due to the possibility to easily model the system evolution using the positive-$\mathcal{P}$ phase-space method, it is possible to treat even large waveguide arrays exactly, which would be exponentially difficult when using the standard methods of calculations in the Fock space or other approximate methods. Such favourable scaling and noiseless operation mirrors that seen in positive-$\mathcal{P}$ treatments of Gaussian boson sampling setups \cite{Drummond22,Dellios23}.

\begin{figure}[bht!]
    \centering
    \includegraphics[width=0.94\linewidth]{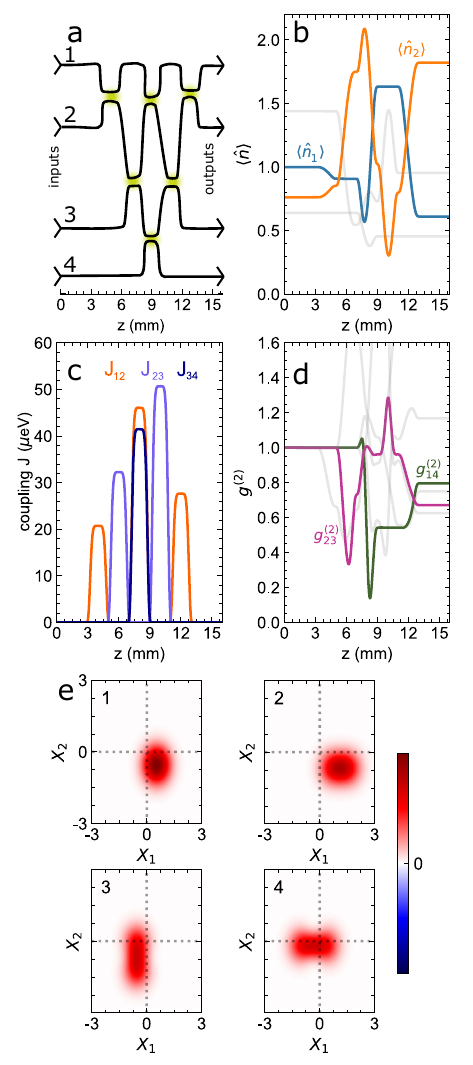}
    \caption{The figure presents a 4-mode waveguide interferometer with the network scheme depicted in panel \textbf{a} and the corresponding coupling strengths shown in panel \textbf{c}. Panels \textbf{b} and \textbf{d} show, respectively, the average photon number in different waveguides and the different $g^{(2)}$ cross-correlation functions. In both panels, two functions are emphasized to demonstrate the dynamics while ensuring readability. Panel \textbf{e} shows the single-mode reduced Wigner functions of all of the modes at the output of the interferometer. The input consists of coherent states at modes 1, 3 and 4 with amplitudes $\beta_{1} = e^{\mathrm{i} \pi/4}$, $\beta_{3} = 1.2\mathrm{i}$, and $\beta_{4} = 0.8e^{\mathrm{i} \pi5/4}$, respectively. Mode 2 input is a kitten state $(1/\mathcal{N})(\ket{\beta} + \ket{-\beta})$ with amplitude $\beta = 1$.}
    \label{fig:fig5}
\end{figure}

\subsection{Binary classification with a four-mode interferometer}

In this section we will study the utilization of a four-mode waveguide interferometer to perform a classification task. The idea is, that in three of the four modes, initially coherent states will propagate, and the data that is classified will be encoded in the phases of two of these coherent state complex amplitudes. The third coherent state will be used for phase reference and feature space expansion, meaning increasing the number of measurable correlations between interferometer outputs. Finally, one of the modes will initially be in a kitten state, providing non-classical correlations to the circuit, which prove to be crucial in performing the classification task, see Fig.~\ref{fig:fig1}.

\begin{figure*}[htb!]
    \centering
    \includegraphics[width=0.9\linewidth]{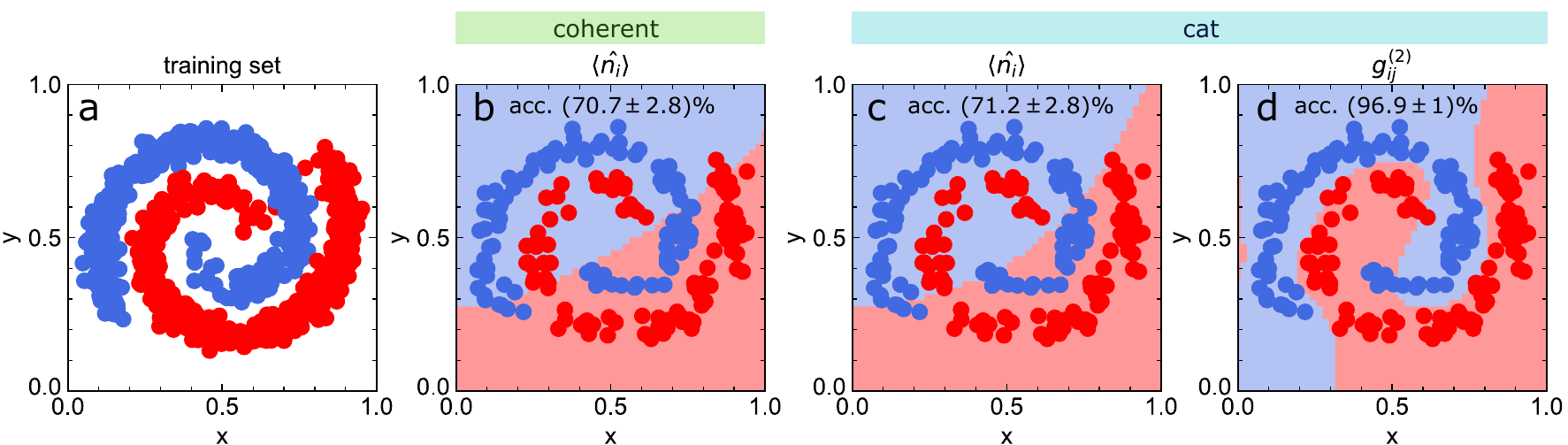}
    \caption{Binary classification on a spiral dataset using the 4-mode linear interferometer of Fig.~\ref{fig:fig5}. Panel \textbf{a} presents the training dataset consisting of two spirals belonging to two distinct classes. The average occupations on each mode, together with all six second-order cross-correlation functions at the output of the interferometer are collected and used for training a logistic regression model. Panel \textbf{b} presents the decision boundary for a model  using only 
    the average output occupations as output features, with all of the interferometer inputs being coherent states.  Panel \textbf{c} presents the decision boundary for a model  using again only 
    the average output occupations as output features, but for which 
    the input of one of the modes (mode 2) was a kitten state. Panel \textbf{d}  presents the decision boundary for a model  using only the six multimode $g^{(2)}$ correlations with a kitten state input in mode2.}
    \label{fig:fig6}
\end{figure*}

We have chosen the geometry of couplings used in the four-mode circuit in such a way, that each state has at least one time interval of interaction 
with every other mode. Graphically, the scheme of the waveguide interferometer and the corresponding coupling profiles are depicted in Fig.~\ref{fig:fig5}\textbf{a} and \textbf{c}, respectively. At the output of the interferometer, the average occupations of all the modes $\langle \hat{n} \rangle$ and different $g^{(2)}$ cross-correlation functions are measured. In total, for a four mode network, the output consists of four average occupations $\left\{\langle \hat{n}_{1} \rangle, \langle \hat{n}_{2} \rangle, \langle \hat{n}_{3} \rangle, \langle \hat{n}_{3} \rangle\right\}$ and six two-photon cross-correlation functions $\left\{ g_{12}^{(2)}, g_{13}^{(2)}, g_{14}^{(2)}, g_{23}^{(2)}, g_{24}^{(2)}, g_{34}^{(2)} \right\}$. In initial research, 
we also considered the local $g^{(2)}$ functions. However, the six cross-correlation functions were sufficient to solve the task. This is a point at which 
we can see the difference between having a non-classical state in the circuit and if we were to use only coherent states. In the latter case, all cross-correlation functions would be equal to one, whereas in the case of a non-classical state the correlation function dynamics is very rich, as exemplified in Fig.~\ref{fig:fig5}\textbf{d}. The output state of the circuit can in principle be a complicated entangled state. Figure \ref{fig:fig5}\textbf{e} shows single-mode Wigner functions at the output of the waveguide array. They 
are not coherent states 
and hence 
adding the kitten state to the network yields interesting evolution of observables as well as nontrivial states at all outputs of the interferometer. 

The dataset that we used to perform classification is the spiral dataset presented in Fig.~\ref{fig:fig6}\textbf{a}, consisting of two interlaced spirals, each colored differently. Such data is not linearly separable and linear algorithms such as pure logistic regression fail to correctly perform the classification. To obtain high accuracy, one needs to use a nonlinear machine learning approach (for example, use an artificial neural network with nonlinear activation functions) or, like we propose in this work, use a linear optical device and measure nonlinear observables.

In order to perform the classification, the $x$ and $y$ point coordinates, which are the input features of this dataset, are encoded in the phase of coherent laser fields. These enter the waveguide interferometer in mode 1 and 3 respectively. Mode 2 has a kitten state as the input, and mode 4 a coherent state with fixed phase. The input states can be written as $\{ \mathrm{mode1:} \ket{\beta = e^{\mathrm{i}\pi x/2}}, \mathrm{mode2:} (1/\mathcal{N})(\ket{\beta = 1} + \ket{\beta = -1}), \mathrm{mode3:} \ket{\beta= 1.4e^{\mathrm{i}(\pi/2 + \pi y/2)}}, \mathrm{mode4:} \ket{\beta = \mathrm{i}}\}$. For each pair ($x$,$y$) we obtain 4 average occupations and 6 cross-correlations at 
the output of the interferometer. Both the average occupation and correlations are nonlinear functions of the initial data features (x and y coordinates). With this nonlinearity, together with the increase of the number of features (by measuring 6 cross-correlation functions), we can treat the interferometer as performing a nonlinear feature space expansion. Finally, via supervised learning we train a software logistic regression model to classify data points based on the outputs of the interferometer, as shown in 
Fig.~\ref{fig:fig1}. 

We chose 
1000 random 400-point subsets out of a computed 800-point dataset for training and testing, with 300 samples chosen as training data and 100 samples used for testing, ensuring that both sets contained equal numbers of elements from each class. 
Results are shown in Fig.~\ref{fig:fig6}.
Background colors represent the calculated decision boundaries.
As seen 
in Fig.~\ref{fig:fig6}\textbf{b,c}, using only average occupations failed to classify the data. This was the case when all input states were coherent (mode 2 was in a coherent state $\ket{\beta = 1}$ instead of a kitten state) with the accuracy being $(70.7 \pm 2.8)$\%, as well as when having a quantum kitten state at the input with 
accuracy 
$(71.2 \pm 2.8)$\%. However, using a vector of cross-correlation functions (this could only be done using a kitten state, since 
for coherent inputs all correlation functions are equal to 1), we  found the accuracy to be significantly higher and equal to $(96.9 \pm 1)$\%.  From the highly nonlinear shape of the decision boundary it is clear that the propagation through the waveguide interferometer, together with phase encoding and correlation measurements is a nonlinear phase space expansion sufficient to classify the dataset. 

\section{Discussion}
The utilization of a linear optical waveguide interferometer to perform a quantum-enhanced machine learning task demonstrates how a simple physical system, where data is encoded as parameters of classical states can be used to solve complex tasks as long as an interaction with a non-classical state is allowed. This interaction introduces correlations between the waveguide modes, which would be absent if all the inputs were coherent states. In principle, the non-classical state may be any superposition of coherent states. We suspect that utilizing single-photon (Fock) states or squeezed coherent states would lead to similar complex dynamics of the occupations and cross-correlations due to their non-classicality. In this work, in order to keep the results exact and show a practical implementation of the generalized-$\mathcal{P}$ distribution, we considered a kitten state as the quantum state injected in the interferometer. 

An essential feature of the setup is the linear scaling of the model with system size $N$ -- the number of variables to simulate grows proportionally to the number of waveguides. Actual computational time needed may scale faster, polynomially, in $N$ if the number of interactions to implement or the evolution time needed depend also on $N$, but still far from the exponential scaling of brute force methods.

Additionally, the simulation of an arbitrary superposition of coherent states in the weighted generalized-$\mathcal{P}$ representation is convenient, as the input state is modeled exactly, whereas in the Fock state basis, there is always a nonzero level of approximation. Although for any quantum state there exist positive-$\mathcal{P}$ probabilistic representations, a constructive prescription such as the canonical one \cite{Drummond_1980,Braunstein} or later findings \cite{Olsen2009} is often quite  
broad and sampling from it is inefficient in comparison to deterministically choosing the inputs using the described generalized-$\mathcal{P}$ representation -- when possible, as with cat states here. 

The physical system we study here is linear, hence the generalized-$\mathcal{P}$ evolution equations are deterministic. However, if there would be a source of nonlinearity, e.g. Kerr ($\hat{a}_{i}^{\dagger}\hat{a}_{i}^{\dagger}\hat{a}_{i}\hat{a}_{i}$) or cross-Kerr ($\hat{a}_{i}^{\dagger}\hat{a}_{i}\hat{a}_{j}^{\dagger}\hat{a}_{j}$) nonlinearity, the evolution equations would become stochastic, and remain stable with sufficient dissipation \cite{Deuar_2021} or short enough times \cite{Deuar06a}. Given a correct preparation of the stochastic noise such a system would be easy to simulate. It would require multiple trajectories to realize the quantum noise present in the evolution equations, but the ensemble size is still reduced with the form (\ref{eq:catP}) compared to canonical-form initial distributions 
since the input conditions can be chosen deterministically.

\section{Conclusions}

In conclusion, this work provides a theoretical analysis of employing a superposition of coherent states to solve a machine learning classification task. We study the generalized-$\mathcal{P}$ method as an ideal framework to model the dynamics of coherent states and their superpositions propagating through a linear optical waveguide interferometer and compare the results to the standard method of solving the von Neumann equation in the Fock state basis. We also show how positive-$\mathcal{P}$ trajectories can
be used to calculate the Wigner function and conveniently visualise the state  by building it out of ``partial'' Wigner functions corresponding to each positive-$\mathcal{P}$ sample.
We study how a coherent state superposition can propagate through the waveguide array and how the average occupations and two-photon cross-correlation functions between the modes evolve throughout the interferometer. Finally, we showed that the collection of observables consisting of average cross-correlation functions can be used to perform a nonlinear feature space expansion where the input data features are encoded in the phases of input classical states. Promising results, with near-perfect ($\sim$ 97\%) classification accuracy with only 4 outputs but depending entirely on the input of a nonclassical state provide a motivation for verifying this theoretical proposal experimentally in future work, and identifying the necessary conditions for quantum advantage with the help of scalable simulation methods presented here.

\section*{methods}
\subsection*{Observables in the Positive-$\mathcal{P}$ framework}
In this section, we provide a step-by-step derivation of the expected value of the number operator\cite{Deuar_2021} and the expected value of measuring $n$ photons in a given quantum state in the Positive-P framework. The expected values of the number operator $\hat{a}^\dagger\hat{a}$ and the operator $\hat{a}^\dagger\hat{a}^\dagger\hat{a}\hat{a}$ (similar derivation) were used to calculate average mode occupations and cross-correlation functions. The expected value of the $n$-th photon state operator $\ket{n}\bra{n}$ was used to find the photon-number distributions presented in Fig.~\ref{fig:fig2} \textbf{a - d}.

The number operator expectation value is well known and can be derived via
\begin{align}
\langle\hat{n}\rangle 
= \langle \hat{a}^\dagger \hat{a} \rangle 
    &= \mathrm{Tr}(\hat{a}^\dagger \hat{a}\,\hat{\rho}) \nonumber \\
    &= \int {d^2}{\alpha}\,{d^2}\tilde{{\alpha}}\,\mathcal{P}({\alpha},{\tilde{\alpha}^*})\, \mathrm{Tr}(\hat{a}^\dagger \hat{a}\, \hat{\Lambda})  \nonumber\\
    &= \int {d^2}{\alpha}\,{d^2}\tilde{{\alpha}}\,\mathcal{P}({\alpha},{\tilde{\alpha}}^*) \alpha 
        \left[ \tilde{\alpha}^{*} + \frac{\partial}{\partial \alpha} \right]
        \mathrm{Tr}(\hat{\Lambda})  \nonumber\\
    &= \int {d^2}{\alpha}\,{d^2}\tilde{{\alpha}}\,\mathcal{P}({\alpha},{\tilde{\alpha}}^*) \alpha\, \tilde{\alpha}^{*}  \nonumber\\
    &= \lim_{S\to\infty} \langle \alpha \tilde{\alpha}^{*} \rangle_S .
\end{align}
The transition from the second to third line of the equation is done by using known derivative identities on $\hat{\Lambda}$  \cite{Drummond_1980,Deuar_2021} such as $\hat{a}\hat{\Lambda}=\alpha\hat{\Lambda}$ and $\hat{a}^{\dagger}\hat{\Lambda}=[\tilde{\alpha}^*+\partial/\partial\alpha]\hat{\Lambda}$ to replace
the expression $\hat{a}^{\dagger} \hat{a} \hat{\Lambda}$. 
The c-numbers and c-number derivatives can later be taken out of the trace. The step from line three to four uses the fact that the trace of the kernel operator is equal to one, and the transition from line four to five makes use of the fact that $\mathcal{P}$ is a positive and real distribution that can be sampled.

The ensemble estimator for the $n$-th Fock state population 
is, apparently, not well known. It can, however, be readily derived:
\begin{align}
\bra{n}\hat{\rho}\ket{n}
  &= \mathrm{Tr}\!\big(\,|n\rangle\langle n|\,\hat{\rho}\,\big) \nonumber \\[6pt]
  &= \int d^2\alpha\,d^2\tilde{\alpha}\;
     \mathcal{P}(\alpha,\tilde{\alpha}^*)\,
     \mathrm{Tr}\!\big(\,|n\rangle\langle n|\,\hat{\Lambda}(\alpha,\tilde{\alpha}^*)\,\big)  \nonumber\\[6pt]
  &= \int d^2\alpha\,d^2\tilde{\alpha}\;
     \mathcal{P}(\alpha,\tilde{\alpha}^*)\,
     \langle n|\hat{\Lambda}(\alpha,\tilde{\alpha}^*)|n\rangle .
\end{align}
Now 
\begin{align}
\langle n|\alpha\rangle &= e^{-|\alpha|^2/2}\,\frac{\alpha^{n}}{\sqrt{n!}}, 
\qquad
\langle\tilde{\alpha}|n\rangle = e^{-|\tilde{\alpha}|^2/2}\,\frac{(\tilde{\alpha}^{*})^{\,n}}{\sqrt{n!}}, \\[6pt]
\langle\tilde{\alpha}|\alpha\rangle &= \exp\!\bigg(-\frac{|\alpha|^2+|\tilde{\alpha}|^2}{2}+\tilde{\alpha}^*\,\alpha\bigg).
\end{align}
and with Eq.~\ref{lambda} therefore
\begin{align}
\langle n|\hat{\Lambda}|n\rangle
&= \frac{\langle n|\alpha\rangle\;\langle\tilde{\alpha}|n\rangle}{\langle\tilde{\alpha}|\alpha\rangle} \nonumber \\[6pt]
&= \frac{ e^{-(|\alpha|^2+|\tilde{\alpha}|^2)/2}\,\dfrac{\alpha^{n}(\tilde{\alpha}^*)^{\,n}}{n!} }
        { \exp\!\big(-\tfrac{|\alpha|^2+|\tilde{\alpha}|^2}{2}+\tilde{\alpha}^*\,\alpha\big) } \nonumber \\[6pt]
&= \frac{\alpha^{n}(\tilde{\alpha}^*)^{\,n}}{n!}\;e^{-\tilde{\alpha}^*\,\alpha}.
\end{align}

\begin{align}
\langle n|\,\hat{\rho}\,|n\rangle
  &= \int d^2\alpha\,d^2\tilde{\alpha}\;
     \mathcal{P}(\alpha,\tilde{\alpha}^*)\,
     \frac{\alpha^n(\tilde{\alpha}^*)^{\,n}}{n!}\,e^{-(\tilde{\alpha}^*)\alpha} \nonumber \\[6pt]
  &= \lim_{S\to\infty}\Big\langle \frac{(\alpha(\tilde{\alpha}^*))^{\,n}}{n!}\,e^{-\tilde{\alpha}^*\alpha}\Big\rangle_S .
\end{align}

\subsection*{Wigner function reconstruction from PP samples}
\label{Wigner}
The Positive-P distribution \( \mathcal{P}(\alpha, \tilde{\alpha}^*) \) satisfies by definition
\begin{equation}
\hat{\rho}
    = \int d^{2}\alpha\, d^{2}\tilde{\alpha}\ 
    P(\alpha, \tilde{\alpha}^*)
    \frac{|\alpha\rangle \langle\tilde{\alpha}|}{\langle \tilde{\alpha} | \alpha \rangle}.
\label{eq:rhoposP}
\end{equation}
Estimating the distribution with $S$ samples as per the usual numerical implementation
approach we sample \(\mathcal{P}\) with pairs
\((\alpha_j, \tilde{\alpha}_j)\), 
letting us write
\begin{equation}
\mathcal{P} \approx \frac{1}{S} \sum_{j=1}^S
\delta(\alpha - \alpha_j) 
\delta(\tilde{\alpha}^* - \tilde{\alpha}^*_j),
\end{equation}
which becomes exact as \( S\to\infty \).  
Substituting into \eqref{eq:rhoposP},
\begin{equation}
\hat{\rho} = 
\frac{1}{S} \sum_j 
\frac{|\alpha_j\rangle\langle\tilde{\alpha}_j|}
     {\langle\tilde{\alpha}_j|\alpha_j\rangle}  \nonumber
\end{equation}
\begin{equation}
= \frac{1}{S}\sum_j 
\exp\!\left(
    -\tilde{\alpha}^*_j \alpha_j 
    + \frac{|\alpha_j|^2}{2}
    + \frac{|\tilde{\alpha}_j|^2}{2}
\right)
|\alpha_j\rangle \langle \tilde{\alpha}_j|  \nonumber
\end{equation}
\begin{equation}
\equiv \frac{1}{S}\sum_j \hat{\rho}_j .
\end{equation}

The operators \(\hat{\rho}_j\) are not necessarily Hermitian or positive-semidefinite, but they live in the same vector space and have unit trace.  
Hermiticity can be enforced by noting that if \((\alpha_j,\tilde{\alpha}_j)\) is sampled, then by symmetry the pair \((\tilde{\alpha}_j,\alpha_j)\) is equally probable.  
Define
\begin{equation}
\bar{\rho}_j = \frac{|\tilde{\alpha}_j\rangle\langle\alpha_j|}{\langle \alpha_j | \tilde{\alpha}_j\rangle}.
\label{dualrho}
\end{equation}
Then
\begin{equation}
\hat{\rho} = \frac{1}{S}\sum_j \frac{\hat{\rho}_j + \bar{\rho}_j}{2}.
\end{equation}

We now reconstruct the Wigner function.  
For a single trajectory (drop index \(j\)), the characteristic function is 
\begin{align}
\chi_W(\lambda,\lambda^*) 
&= \mathrm{Tr}\!\left\{ 
    \hat{\rho}\, e^{\lambda \hat{a}^\dagger - \lambda^*\hat{a}}
\right\} \nonumber \\
&= \mathrm{Tr}\!\left\{
    e^{\lambda\hat{a}^\dagger - \lambda^*\hat{a}} \hat{\rho}
\right\} \nonumber \\
&= \mathrm{Tr}\!\left\{
    \hat{D}(\lambda) |\alpha\rangle\langle\tilde{\alpha}|
\right\} 
/ \langle\tilde{\alpha}|\alpha\rangle  \nonumber\\
&= \mathrm{Tr}\!\left\{
    \hat{D}(\lambda)\hat{D}(\alpha)|0\rangle
    \langle\tilde{\alpha}|
\right\}
/ \langle\tilde{\alpha}|\alpha\rangle \nonumber \\
&= e^{(\lambda\alpha^* - \lambda^*\alpha)/2}
     \mathrm{Tr}\!\left\{
         \hat{D}(\lambda + \alpha)|0\rangle\langle\tilde{\alpha}|
     \right\}
/ \langle\tilde{\alpha}|\alpha\rangle \nonumber \\
&= e^{(\lambda\alpha^* - \lambda^*\alpha)/2}
    \frac{\langle\tilde{\alpha}|\lambda+\alpha\rangle}
         {\langle\tilde{\alpha}|\alpha\rangle} \nonumber \\
&= \exp\!\left(
    \tilde{\alpha}^*\lambda
    - \alpha\lambda^*
    - \frac{|\lambda|^2}{2}
\right),
\end{align}
where the displacement operator is given by
\begin{equation}
    \hat{D}(\alpha) = \exp\left(\alpha \hat{a}^\dagger - \alpha^* \hat{a}\right).
\end{equation}

The Wigner function is
\begin{equation}
W_j(\xi,\xi^*) 
    = \frac{1}{\pi^2} 
      \iint d\lambda\, d\lambda^*\,
      e^{-\lambda\xi^* + \lambda^*\xi}
      \chi_W(\lambda,\lambda^*).
\end{equation}
Substituting,
\begin{align}
&W_j(\xi,\xi^*) =
\frac{1}{\pi^2}
\iint d\lambda\, d\lambda^*
\exp\!\big[ \nonumber \\
&-\lambda(\xi^* - \tilde{\alpha}^*) + \lambda^*(\xi - \alpha) - \frac{|\lambda|^2}{2}
\big].
\end{align}

Splitting into real and imaginary integrals, both Gaussian,
\begin{equation}
W_j(\xi,\xi^*) =
\frac{2}{\pi}
\exp\!\left[ -2(\xi^* - \tilde{\alpha}^*)(\xi - \alpha) \right].
\end{equation}

If \(\tilde{\alpha}^* = \alpha^*\), this reduces to the Wigner function of a coherent state.  
In general the expression is not real; symmetrizing with exchanged \(\alpha\leftrightarrow\tilde{\alpha}\) as per (\ref{dualrho}) yields
\[
W_j^{(\mathrm{Herm})} = \mathrm{Re}\{W_j\}.
\]

Finally, summing over $j$, the full Wigner function is
\begin{equation}
W(\xi,\xi^*)
    = \lim_{S\to\infty}\frac{2}{S\pi}
      \sum_{j=1}^S
      \mathrm{Re}\left[
          \exp\!\left(
              -2(\xi^* - \tilde{\alpha}_j^{*})
               (\xi - \alpha_j)
          \right)
      \right].\label{W1}
\end{equation}

This is true for a positive-$\mathcal{P}$ distribution, for a generalized-$\mathcal{P}$ distribution the final result needs to be altered, which can intuitively be understood as multiplying each trajectory by it's corresponding weight. The weight (can be complex) is the same $\mathrm{w}_{\nu}$ for all of the trajectories sampled from a given $\mathcal{P}_{\nu}^+$, producing a $W_{\nu}$ as per (\ref{W1}). 
Applying straight multiplication by $\textrm{w}_{\nu}$ to each weighted contribution $W_{\nu}$ leads in general to a complex function because partial $\mathcal{P}^+_{\nu}$ need not be identical under the $\alpha\leftrightarrow\tilde{\alpha}$ swap. As a result of the partial density matrix contributions not necessarily being Hermitian. However, knowing that the underlying \textit{full} density matrix $\hat{\rho}$ must be Hermitian, each complex weighted contribution to it must have a contributing Hermitian conjugate. Therefore, even though $\mathrm{w}_{\nu}\mathcal{P}^+_{\nu}(\alpha,\tilde{\alpha}) \neq \mathrm{w}^*_{\nu}\mathcal{P}^+_{\nu}(\tilde{\alpha},\alpha)$ in general, a sample pair $\alpha_j,\tilde{\alpha}_j$ with weight $\mathrm{w}$ does have an equally probable pair $\tilde{\alpha}_j,\alpha_j$ pair with weight $\mathrm{w}'=\mathrm{w}^*$ in a possibly different partial distribution. Hence by symmetry we can write

\begin{align}
W(\xi,\xi^*) &= \sum_{\nu}\lim_{S\to\infty}\frac{2}{S\pi}
\sum_{j=1}^S \Re\Big\{ \nonumber \\
&\mathrm{w}_{\nu}\,
\exp\big(-2(\xi^*-\tilde\alpha_{\nu,j}^*)(\xi-\alpha_{\nu,j})\big)\Big\}.\label{Ww}
\end{align}

\subsection*{Post-processing and training}
Once the training and testing datasets are built from the outputs of the quantum network, in order to perform classification using logistic regression in the most optimal manner, it is necessary to normalize the data. This is done by calculating the mean and standard deviation of the training dataset and rescaling both the training and testing data by removing the training data mean and dividing by the standard deviation. For the training of the classification model we have used logistic regression with the Limited-memory BFGS model implemented in the \textit{Scikit Learn} library, which is the algorithm of choice for logistic regression. During training, the minimized loss function is the logistic loss function
\begin{align}
    \ell(\mathcal{W},b) &= - \sum_{i} \bigl[ y_i \log p_i + (1 - y_i) \log (1 - p_i) \bigr], \\
    &\text{where     } p_i = \sigma(\mathcal{W}^\top x_i + b), \: \sigma(z) = \frac{1}{1 + e^{-z}},
\end{align}
$\mathcal{W}$ is the optimized weight matrix, $b$ is the optimized bias vector, $x_{i}$ are the data points and $y_{i}$ are the correct data labels.
The model supports L2 regularization. However, we found via a grid search algorithm over many L2 regularisation strengths that for our training dataset, where the number of features was limited to 4 (for training on occupations) or 6 (for training on correlations) per training example, the model converged best when having no regularization.

\section*{Acknowledgments}
AO acknowledges support from the National Science Center, Poland (PL), Grant No. 2024/52/C/ST3/00324.
This work was financed
by the European Union EIC Pathfinder Challenges
project ``Quantum Optical Networks based on Exciton-polaritons'' (Q-ONE, Id: 101115575) and ``QUantum reservoir cOmputing based on eNgineered DEfect NetworkS in trAnsition meTal dichalcogEnides'' (QUONDENSATE, Id: 101130384). The Center for Quantum-Enabled Computing project is carried out within the International Research Agendas programme of the Foundation for Polish Science co-financed by the European Union under the European Funds for Smart Economy 2021-2027 (FENG).

\section*{Data Availability}

The data that support the findings of this article are openly available at \cite{dane}.

\section*{Code Availability}

The code used for simulating the reservoir dynamics is available from the authors upon  request.

\bibliography{bib_1_}
\end{document}